\documentclass[10pt]{article}
\usepackage[utf8]{inputenc}
\usepackage[super,numbers,sort&compress]{natbib}
\usepackage{authblk}
\usepackage{amsmath,physics,bm}
\usepackage{graphicx}
\usepackage[top=1 in,bottom=1in, left=1 in, right=1 in]{geometry}
\usepackage{setspace}

\title{\Large{\bf{Spin Hall effect of light under arbitrarily polarized and unpolarized light}}}
\author[1]{Minkyung Kim}
\author[1]{Dasol Lee}
\author[1,2,*]{Junsuk Rho}
\affil[1]{\small{Department of Mechanical Engineering, Pohang University of Science and Technology (POSTECH), Pohang 37673, Republic of Korea}}
\affil[2]{\small{Department of Chemical Engineering, Pohang University of Science and Technology (POSTECH), Pohang 37673, Republic of Korea}}
\affil[*]{\small{jsrho@postech.ac.kr}}
\date{}

\renewcommand{\vec}[1]{\mathbf{#1}}

\begin{document}

\maketitle

\begin{abstract}
The spin Hall effect of light (SHEL), which refers to a spin-dependent and transverse splitting at refraction and reflection phenomena, inherently depends on the polarization states of the incidence. Most of the previous research have focused on a horizontally or vertically polarized incidence, in which the analytic formula of the shift is well-formulated and SHEL appears symmetrically in both shift and intensity. However, the SHEL under an arbitrarily polarized or unpolarized incidence has remained largely unexplored. Whereas the SHEL under other polarization is sensitive to incident polarization and is asymmetrical, here we demonstrate that the SHEL is independent of the incident polarization and is symmetrical in shift if Fresnel coefficients of the two linear polarization are the same. The independence of the shift with respect to the incident polarization is proved both analytically and numerically. Moreover, we prove that under an unpolarized incidence composed of a large number of completely random polarization states, the reflected beam is split in half into two circularly polarized components that undergo the same amount of splitting but in opposite directions. This result means that under unpolarized incidence, the SHEL occurs exactly the same as under horizontally or vertically polarized incidence. We believe that the incident-polarization-independent SHEL can broaden the applicability of the SHEL to cover optical systems in which polarization is ill-defined.
\end{abstract}

\section{Introduction}
Spin-orbit coupling is a universal phenomenon that can be observed in a variety of fields in physics including classical mechanics \cite{coriolis}, quantum physics \cite{thomas1926motion}, and photonics \cite{bliokh2015spin}. In particular, there exist diverse kinds of spin-orbit interactions in photonics because of rich physics enabled by the two electric and magnetic vector fields \cite{born2013principles}. Among them, one interesting spin-orbit related feature is the spin Hall effect of light (SHEL) \cite{onoda2004hall, hosten2008observation, Ling_2017}, also known as an Imbert-Fedorov shift \cite{fedorov2013theory, PhysRevD.5.787}, which reveals itself as a transverse and spin-dependent splitting of a finitely-thick beam at an optical interface. A physical mechanism that underpins the spin-dependent shift is the transverse nature of light, $\vec{k} \cdot \vec{E} = 0$, that makes the incidence contain a bundle of slightly differently defined polarization bases \cite{Bliokh_2013}. 

Despite its long history tracing back to the mid-19th century, the SHEL has regained a booming interest recently especially in photonics and metamaterials communities. A variety of nanophotonic devices and metamaterials have been proposed to enlarge the spin-dependent shift \cite{yin2013photonic, Takayama:18, kim2019observation, doi:10.1063/5.0009616, Zhu:15, Jiang_2018_BIC, PhysRevLett.124.053902, 7394110}, to increase the efficiency \cite{https://doi.org/10.1002/lpor.202000393}, and to exploit the SHEL to identify geometric parameters \cite{wang2020probing, doi:10.1063/1.4772502}, electric and magnetic properties \cite{PhysRevApplied.13.014057, Li:20}, and chemical reactions \cite{doi:10.1063/1.5131183, doi:10.1063/1.5130729} with high precision. Except for the studies of asymmetric SHEL, in which the magnitude of the shift is spin-dependent \cite{Zhou:16, zhu2017upper, Jiang:18, Zhou:19, WU2020396}, most previous studies have focused only on horizontally or vertically polarized incidence. By symmetry, a horizontally or vertically polarized light is split at the optical interface into equal amplitudes of left circularly polarized (LCP) and right circularly polarized (RCP) beams, which shift by the same magnitude but with opposite signs \cite{hosten2008observation}. However, under an arbitrary incident polarization, neither the magnitude of the shift nor the intensity of the two circularly polarized components is symmetrical in general; in other words, the arbitrarily polarized incidence is split into LCP and RCP unevenly in both shift and intensity. Furthermore, the shift is sensitive to the polarization states of the incidence, resulting in a different amount of splitting as the incident polarization varies.

Until very recently, a well-defined polarization state has been regarded as a prerequisite of spin-orbit related phenomena. However, it has been reported recently that even unpolarized light can induce a transverse spin \cite{eismann2020transverse}. Here, we demonstrate that if reflection coefficients for $s$ and $p$ polarizations are equal to each other, the shift is degenerate for any arbitrarily polarized incidence and that the SHEL appears symmetrically in both shift and intensity under unpolarized incidence, as it does under a horizontally or vertically polarized incidence. First, we prove both analytically and numerically that the SHEL is independent of the polarization state of the incidence when the two linear polarizations have the same reflection coefficients. In such a case, LCP and RCP components of the reflected beam are shifted evenly, by the same amount but in opposite directions. The intensities of the LCP and RCP components are generally asymmetrical under an arbitrarily given incident polarization, but they also become symmetrical when the incidence is unpolarized, i.e., when the incidence is a superposition state of a vast number of completely random polarizations. We show that when the reflection coefficients are degenerate, the SHEL appears symmetrically in both shift and intensity, thereby having the whole symmetries of the SHEL under horizontally or vertically polarized incidence even under unpolarized incidence. In stark contrast to the previous studies of the SHEL where the incidence has a well-defined single polarization state, our work will extend the field of the SHEL to include unpolarized sources.

\section{Results and Discussion}
\subsection{The analytical proof of incident-polarization-independent shift}
This section is devoted to proving the incident-polarization-independent shift theoretically by using a wave packet model. To do so, we examine how an incidence characterized by a single arbitrary polarization transforms at an interface through the reflection. An incident Gaussian beam propagating in the  $x_I$-$z_I$ plane along the $z_I$-axis can be expressed in momentum space as
\begin{equation}
	\bm{\psi}_I = \begin{pmatrix} \psi_I^H \\ \psi_I^V \end{pmatrix} \psi_0,
	\label{incidence}
\end{equation}
where the superscripts $H$ and $V$ correspond to horizontal and vertical polarization respectively, $\begin{pmatrix} \psi^H_I & \psi^V_I \end{pmatrix}^T$ is the Jones vector of the incidence, and $\psi_0$ is a Gaussian term given as
\begin{equation}
	\psi_0 = \frac{\omega_0}{\sqrt{2\pi}} \exp(-\frac{\omega_0^2}{4}(k_x^2 + k_y^2)),
	\label{psi0}
\end{equation}
where $\omega_0$ is a beam waist, $k_0$ is the incident wave vector, and $k_x \equiv k_{Ix} = -k_{Rx}$ and $k_y \equiv k_{Iy}  = k_{Ry}$ are $x$- and $y$-components of the wave vector. The Jones vectors of the incident and reflected beams represented in a linear basis are related to each other \cite{PhysRevA.84.043806}
\begin{equation}
	\begin{pmatrix} \psi^H_R \\ \psi^V_R \end{pmatrix} = \begin{pmatrix} r_p & \frac{k_y}{k_0} (r_p + r_s) \cot{\theta_i} \\ -\frac{k_y}{k_0} (r_p + r_s) \cot{\theta_i} & r_s \end{pmatrix} \begin{pmatrix} \psi^H_I \\ \psi^V_I \end{pmatrix},
	\label{s1}
\end{equation}
where the subscripts $I$ and $R$ correspond to the incident and reflected beam respectively, $r_s$ and $r_p$ are Fresnel reflection coefficients for $s$ and $p$ polarization, and $\theta_i$ is an incident angle. The reflected beam is composed of the Jones vector and the Gaussian part similarly to Eq. \ref{incidence},
\begin{equation}
	\bm{\psi}_R = \begin{pmatrix} \psi_R^H \\ \psi_R^V \end{pmatrix} \psi_0.
	\label{reflected}
\end{equation}
The reflected beam can be represented in a circular basis by using a basis transformation
\begin{equation}
	\begin{pmatrix} \psi^+_R \\ \psi^-_R \end{pmatrix} = \frac{1}{\sqrt{2}} \begin{pmatrix} 1 & i \\ 1 & -i \end{pmatrix} \begin{pmatrix} \psi^H_R \\ \psi^V_R \end{pmatrix},
	\label{hv2lr}
\end{equation}
where the superscripts $+$ and $-$ denote LCP and RCP respectively. Since the wave vector component along the transverse axis is much smaller than the wave number ($k_y/k_0 \ll 1$), the first-order Taylor expansion of $1 \pm x \approx \exp(\pm x)$ can be applied. Then, Eq. \ref{s1} can be expressed as
\begin{equation}
	\begin{pmatrix} \psi^+_R \\ \psi^-_R \end{pmatrix} = \frac{1}{\sqrt{2}} \begin{pmatrix} r_p \exp (+i k_y \triangle_H ) & i r_s \exp (+i k_y \triangle_V ) \\ r_p \exp (-i k_y \triangle_H) & -i r_s \exp (-i k_y \triangle_V ) \end{pmatrix} \begin{pmatrix} \psi^H_I \\ \psi^V_I \end{pmatrix},
	\label{s2}
\end{equation}
where
\begin{align}
	\triangle_H &= \frac{\cot{\theta_i}}{k_0}(1 + \frac{r_s}{r_p}), \notag \\
	\triangle_V &= \frac{\cot{\theta_i}}{k_0}(1 + \frac{r_p}{r_s}).
	\label{triangle}
\end{align}
Then the shift can be obtained by using a position operator $\vec{r} = i\hbar \partial_\vec{p} = i \partial_\vec{k}$ to calculate an average value of $y$ position of the reflected beam as
\begin{equation}
	\delta^\pm = \text{Re} \frac{\expval{i\partial_{k_y}}{\psi_R^\pm \psi_0}}{\bra{\psi_R^\pm \psi_0}\ket{\psi_R^\pm \psi_0}}.
	\label{shift_pm}
\end{equation}
If the polarization state of the incidence is either horizontal ($\psi_I^H = 1, \psi_I^V = 0$) or vertical ($\psi_I^H = 0, \psi_I^V = 1$), then $\psi_R^\pm$ are eigenstates of the position operator $i\partial_{k_y}$ with eigenvalues of $\mp\triangle_H$ and $\mp\triangle_V$ respectively. On the other hand, $i\partial_{k_y} \psi_0 = -\psi_0 k_y \omega_0^2/2$, and this odd function has no contribution in the indefinite integral. Therefore, the shift can be obtained by taking the eigenvalues of $\psi_R^\pm$, so Eq. \ref{shift_pm} gives the well-known formulas \cite{qin2009measurement, hosten2008observation}
\begin{align}
	\delta_H^\pm &= \mp \frac{\cot{\theta_i}}{k_0} \text{Re} (1 + \frac{r_s}{r_p}), \notag \\
	\delta_V^\pm &= \mp \frac{\cot{\theta_i}}{k_0} \text{Re} (1 + \frac{r_p}{r_s}).
	\label{shift_HV}
\end{align}

Under an incidence that is polarized neither horizontally nor vertically, the shift should be calculated by substituting $\psi_R^\pm$ (Eq. \ref{s2}) into Eq. \ref{shift_pm} and by performing integration in momentum space. Interestingly, when the Fresnel coefficients of the two linear polarizations are equal ($r_s = r_p \equiv r$), the two equations in Eq. \ref{triangle} are degenerate ($\triangle_H = \triangle_V \equiv \triangle$), then $\psi_R^\pm$ become eigenstates of the operator $i\partial_{k_y}$ with eigenvalues of $\mp\triangle$ for any incident polarization. Consequently, similarly to the instance under a horizontally or vertically polarized incidence, Eq. \ref{shift_pm} reduces to $\delta^\pm =  \mp \text{Re}(\triangle)$, which is equivalent to Eq. \ref{shift_HV} but under an arbitrarily polarized incidence. This formula of the shift contains neither $\psi^H_I$ nor $\psi^V_I$ but only depends on $\theta_i$. The independence of $\delta^\pm$ with respect to the incident polarization clearly shows that an arbitrarily polarized incidence is shifted by the same distance, regardless of the polarization states of the incidence. Another important attribute of the SHEL that originates from $r_s = r_p$ is that the splitting occurs symmetrically in shift ($\lvert \delta^+ \rvert = \lvert \delta^- \rvert$) under an arbitrarily polarized incidence. This result is opposed to the conventional wisdom that a circularly or elliptically polarized incidence is split asymmetrically into LCP and RCP \cite{Zhou:16, zhu2017upper, Jiang:18, Zhou:19, WU2020396}.

Although the condition of $r_s = r_p$ seems to imply the polarization-independent reflection, it is not true. Instead, $r_s = r_p$ is associated with the $\pi$ phase shift between the electric fields of $s$- and $p$-polarized reflected beam because of the sign convention \cite{born2013principles}. Polarization-independent reflection ($r_s = -r_p$) is another possible solution of $\triangle_H = \triangle_V$, but is excluded because it leads to zero shift. Here we consider only the reflection type of the SHEL for demonstration but this scheme is also applicable to the transmission type by matching $t_s = -t_p$. 

\subsection{The intensities of circularly polarized components of the reflected beam under arbitrarily polarized incidence}
To understand how the SHEL occurs systematically, not only the shifts but also the intensities of the LCP and RCP components of the reflected beam should be considered. For a given incident polarization, the intensities can be calculated as
\begin{equation}
	I_R^\pm = \frac{\bra{\psi_R^\pm \psi_0}\ket{\psi_R^\pm \psi_0}}{\bra{\psi_I^H \psi_0}\ket{\psi_I^H \psi_0} + \bra{\psi_I^V \psi_0}\ket{\psi_I^V \psi_0}}.
	\label{IR}
\end{equation}
Considering that both $\psi_I^H$ and $\psi_I^V$ are complex, the intensities of the two circularly polarized beams are generally not symmetrical ($I_R^+ \neq I_R^-$). Especially, when $r_s = r_p$, Eq. \ref{IR} can be simplified to
\begin{equation}
	I_R^\pm = \lvert r \rvert ^2 \Big[ \frac{1}{2} \pm \lvert \psi_I^H \rvert \lvert \psi_I^V \rvert \sin \Big(\text{arg}(\psi_I^H) - \text{arg}(\psi_I^V) \Big) \Big].
	\label{IRpm}
\end{equation}
This result shows that even when $r_s = r_p$ is satisfied, $I_R^+ = I_R^-$ if the incidence is linearly polarized, and $I_R^+ \neq I_R^-$ otherwise. It also conforms to our intuition in that only linearly polarized light is a superposition of equal amounts of LCP and RCP whereas the others such as elliptically and circularly polarized light are not.

\subsection{The SHEL under arbitrarily polarized incidence and unpolarized incidence}
\begin{figure}[h!] \centering
	\includegraphics[width = 0.7031\textwidth]{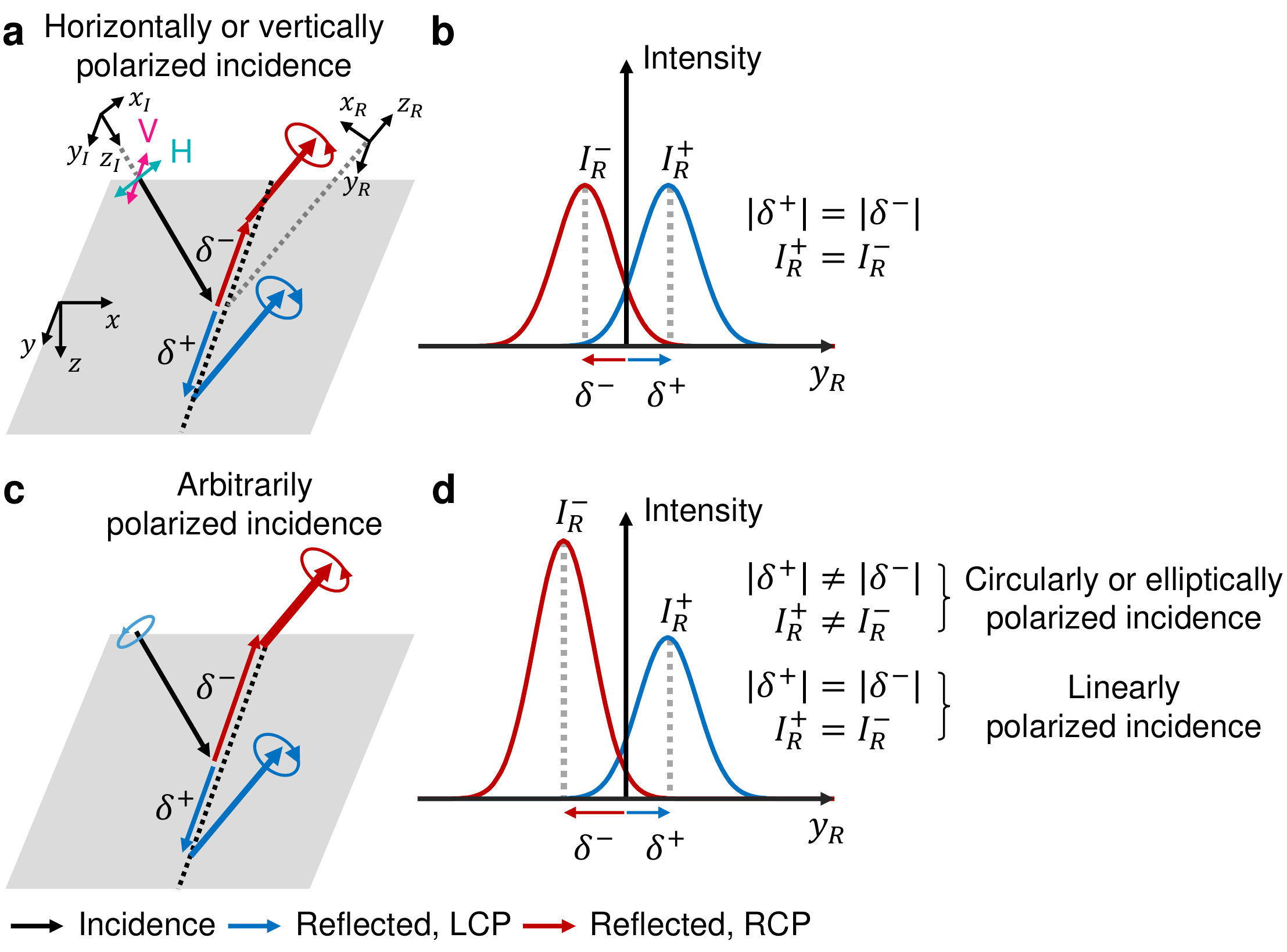} 
	\caption{Schematics of the SHEL under various polarization states of incidence in a general case ($r_s \neq r_p$). (a) The SHEL in real space and (b) corresponding intensity profiles of the reflected beam along the transverse axis when the incidence is horizontally or vertically polarized. The two circularly polarized reflected beams are symmetrical in shift ($\lvert \delta^+ \rvert = \lvert \delta^- \rvert$) and intensity ($I_R^+ = I_R^-$). (c) The SHEL in real space and (d) corresponding intensity profiles of the reflected beam when the incidence has an arbitrary polarization. Both symmetries are preserved under linear polarization but are broken ($\lvert \delta^+ \rvert \neq \lvert \delta^- \rvert$ and $I_R^+ \neq I_R^-$) under circularly or elliptically polarized incidence. For clear visualization, only the central wave vector component of the incident and reflected beams are plotted, instead of the wave packets with finite beam waist. The width of the central wave vectors of the reflected beams shown in Fig. \ref{schematic1}a and c represent their intensities.}
	\label{schematic1}
\end{figure}
Using Eq. \ref{shift_pm} and Eq. \ref{IRpm}, which provide the shift and intensities of the circularly polarized components of the reflected beam respectively, we now illustrate how the SHEL occurs under a given incident polarization. Firstly, a general case of $r_s \neq r_p$ is considered. Under a horizontally or vertically polarized incidence, as is well-known from many previous publications \cite{hosten2008observation, qin2009measurement, zhou2014observation}, the reflected beam is split in half into LCP and RCP ($I_R^+ = I_R^-$), which undergoes the equal amount of the shift along the opposite direction ($\lvert \delta^+ \rvert = \lvert \delta^- \rvert$) (Fig. \ref{schematic1}a and b). This symmetrical splitting originates from the high symmetries of the incident polarization state and maintains under other linear polarization. In contrast, under a circularly or elliptically polarized incidence, the polarization breaks the symmetries and thus both the shift and intensity are asymmetrical ($\lvert \delta^+ \rvert \neq \lvert \delta^- \rvert$ and $I_R^+ \neq I_R^-$) (Fig. \ref{schematic1}c and d).

\begin{figure}[h!] \centering
	\includegraphics[width = 0.7031\textwidth]{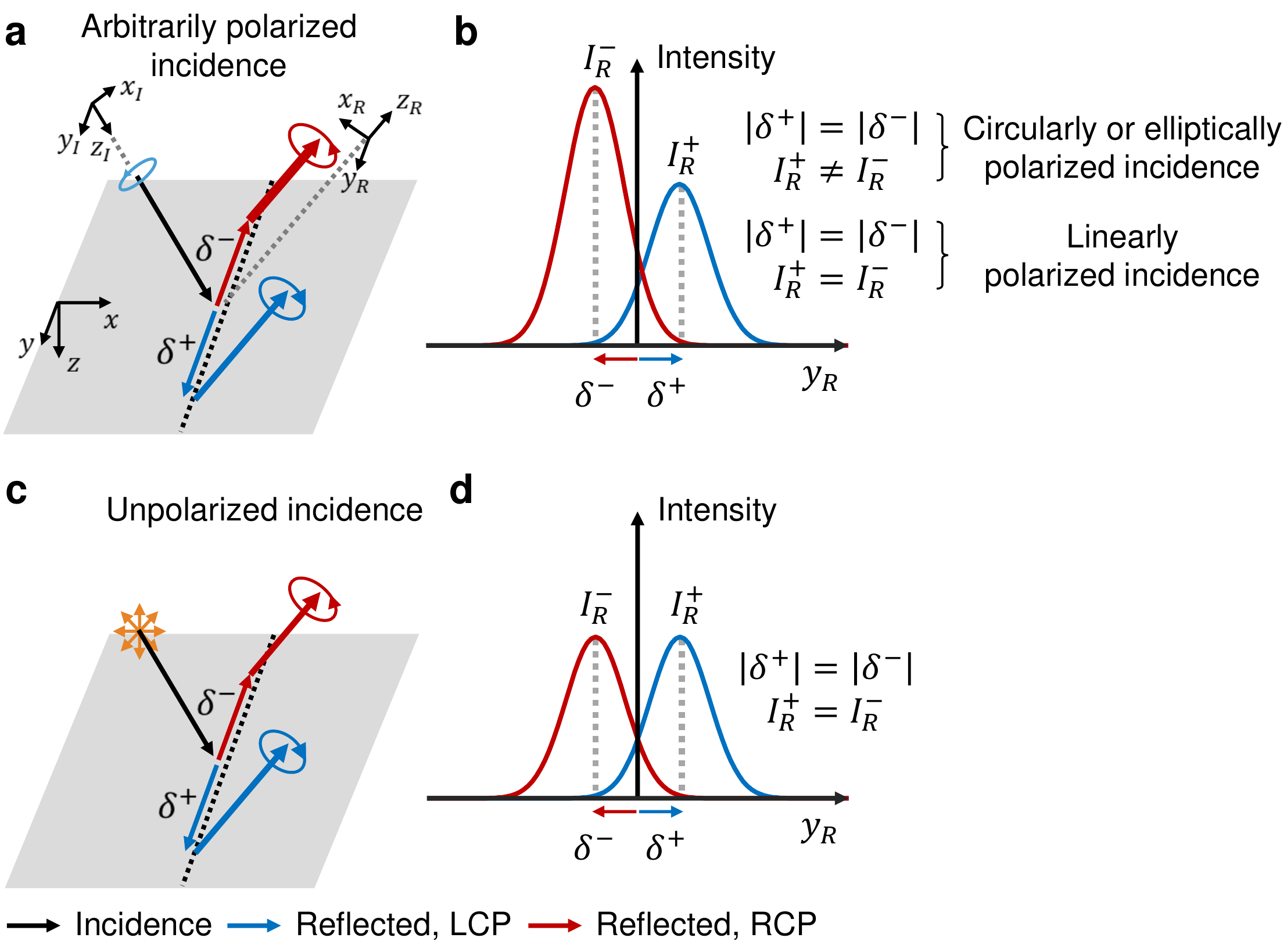} 
	\caption{Schematics of the SHEL for various polarization states of incidence when $r_s = r_p$. (a) The SHEL in real space and (b) corresponding intensity profiles of the reflected beam along the transverse axis under arbitrarily polarized incidence. The shift is incident-polarization-independent and is symmetrical ($\lvert \delta^+ \rvert = \lvert \delta^- \rvert$). Intensity is symmetrical ($I_R^+ = I_R^-$) only under a linear polarization, and is asymmetrical ($I_R^+ \neq I_R^-$) under a circularly or elliptically polarized incidence. (c) The SHEL in real space and (d) corresponding intensity profiles of the reflected beam under unpolarized incidence. Both the shift and intensity are symmetrical ($\lvert \delta^+ \rvert = \lvert \delta^- \rvert$ and $I_R^+ = I_R^-$). For clear visualization, only the central wave vector component of the incident and reflected beams are plotted, instead of the wave packets with finite beam waist. The width of the central wave vectors of the reflected beams shown in Fig. \ref{schematic2}a and c represent their intensities.}
	\label{schematic2}
\end{figure}
In contrast, if $r_s = r_p$, the shift becomes symmetrical ($\lvert \delta^+ \rvert = \lvert \delta^- \rvert$) under arbitrarily polarized incidence (Fig. \ref{schematic2}a and b) as proved in the previous section. More importantly, any incidence is split by the same amount regardless of its polarization state. However, despite the degeneracy in the shift, the intensities of the LCP and RCP components are symmetrical ($I_R^+ = I_R^-$) only under a linear polarization but are polarization-dependent and are asymmetrical ($I_R^+ \neq I_R^-$) in a general circularly or elliptically polarized incidence. The intensities can be made symmetrical by using unpolarized incidence. Here, the unpolarized state refers to a superposition of a vast number of randomly polarized light. Then the total incidence has no phase difference between the two components of the Jones vector due to the randomness and hence the sine term in Eq. \ref{IRpm} averages to zero. Therefore, the unpolarized incidence is split in half into LCP and RCP ($I_R^+ = I_R^-$), the shifts of which have the same magnitude ($\lvert \delta^+ \rvert = \lvert \delta^- \rvert$) just as a horizontally or vertically polarized incidence does (Fig. \ref{schematic2}c and d). 

On the other hand, when $r_s$ and $r_p$ have nonzero phase difference, the SHEL has asymmetrical intensity ($I_R^+ \neq I_R^-$) under unpolarized incidence. This phenomenon occurs because when $r_s \neq r_p$, one can still obtain the formula of the intensities of the LCP and RCP components of the reflected beam akin to Eq. \ref{IRpm}, but the sine term of that formula includes not only the phase difference between $\psi_I^H$ and $\psi_I^V$ but also that between $r_s$ and $r_p$. It prevents $I_R^\pm$ from being degenerate, resulting in asymmetrical intensities under unpolarized incidence.

\subsection{Numerical demonstration of the incident-polarization-independent SHEL}
\begin{figure}[h!] \centering
	\includegraphics[width = 0.9375\textwidth]{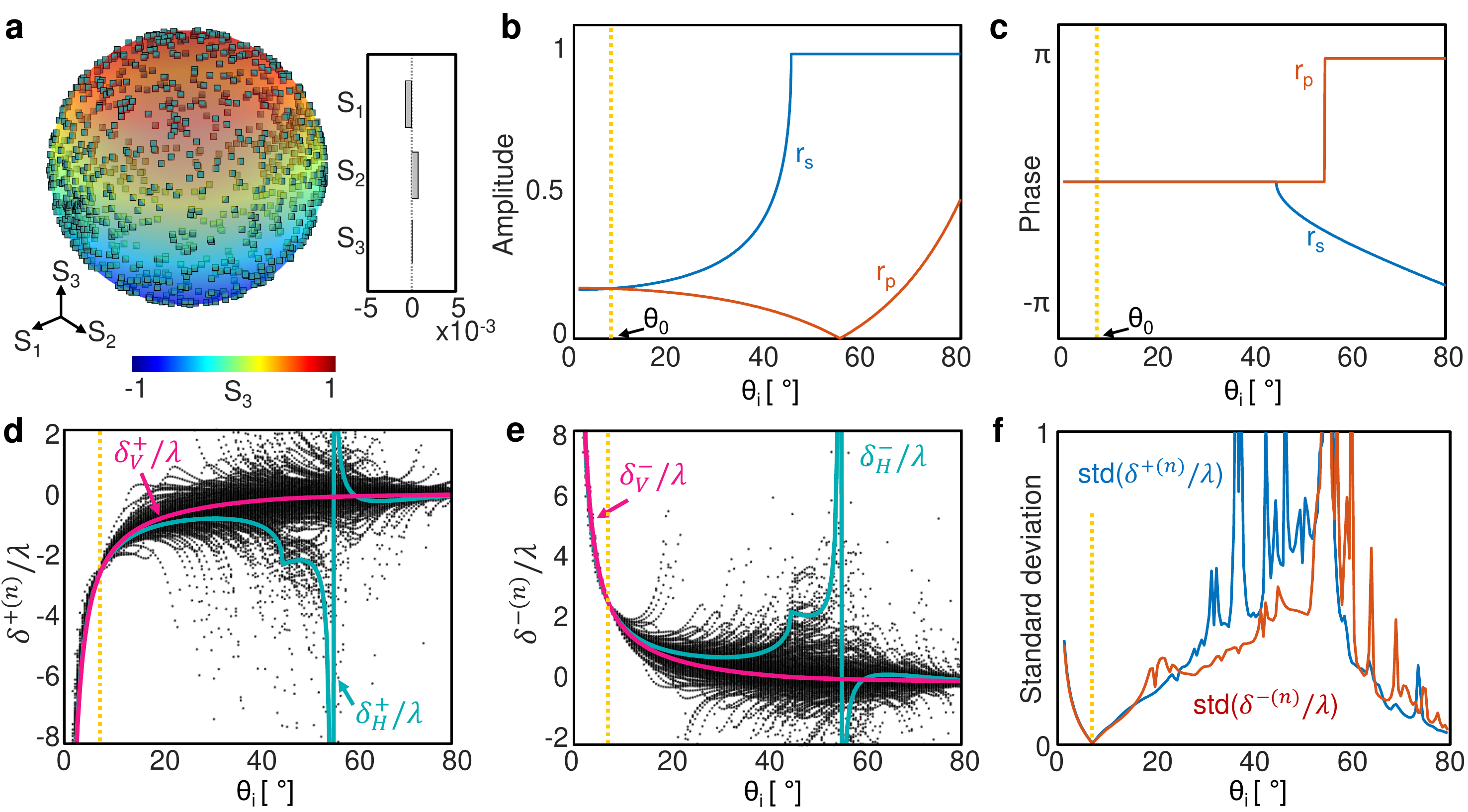} 
	\caption{The SHEL under arbitrarily polarized incidence. (a) Poincar\'e sphere representing $N$ randomly distributed polarization states of incidence (left) and the averaged Stokes parameters (right). (b) Amplitude and (c) phase of $r_s$ and $r_p$ at an interface between an isotropic ($\varepsilon_1 = 2$) and anisotropic ($\varepsilon_{2x} = \varepsilon_{2z} = 4.1, \varepsilon_{2y} = 1$) media. Yellow dashed lines indicate $\theta_0$, an incident angle at which $r_s = r_p$ is satisfied. (d, e) Scatter plots of (d) $\delta^{+(n)}/\lambda$ and (e) $\delta^{-(n)}/\lambda$, where $\lambda$ is the wavelength, under $N$ independent incidences at the interface between the isotropic and anisotropic media. Solid curves represent $\delta_{H, V}^\pm/\lambda$ obtained by Eq. \ref{shift_HV}. (f) Standard deviation of the $N$ values of $\delta^{\pm(n)}/\lambda$ at each $\theta_i$. For simulation, $N = 1000, w_0 = 10^3 \lambda, z_R = 10 \lambda$ are used.}
	\label{random}
\end{figure}

To confirm the incident-polarization-independent SHEL numerically, a Monte Carlo simulation is performed by generating $N$ numbers of randomly oriented polarization states, then using each of them as an incidence. We consider a paraxial regime in which the polarization is defined in a two-dimensional sense. The $n$-th incidence ($n \in \{1, 2, ..., N\}$) has a well-defined polarization that corresponds to the Stokes parameters ($S_1^{(n)}, S_2^{(n)}, S_3^{(n)}$) distributed randomly on a Poincar\'e sphere (Fig. \ref{random}a, left), and have a degree of paraxial two-dimensional polarization equal to unity ($P_{2D}^{(n)} = \sqrt{\sum_{i=1}^3 \big(S_i^{(n)} \big)^2}/S_0^{(n)} = 1$). The averaged Stokes parameters ($\sum_{n=1}^N {S_i^{(n)}/N}$ for $i = 1, 2, 3$) converge to zero as $N$ increases and are all less than $5\times 10^{-3}$ when $N = 1000$ (Fig. \ref{random}a, right). The total incidence satisfies the condition of two-dimensional unpolarized light, that is \cite{eismann2020transverse}, $(S_0, S_1, S_2, S_3) \propto (1, 0, 0, 0)$ and $P_{2D} = 0$. 

To investigate the polarization dependency of the SHEL under arbitrarily polarized incidence, an interface between an isotropic ($\varepsilon_1$) and anisotropic ($\varepsilon_{2x} = \varepsilon_{2z} \neq \varepsilon_{2y}$) media is considered. Whereas $r_s$ and $r_p$ are coupled to each other by Fresnel equations \cite{born2013principles} at an interface between two isotropic media, $s$- and $p$-polarized light experience the anisotropic medium, in which the optic axis lies perpendicular to the incident plane, independently. When $\varepsilon_{2x} = \varepsilon_{2z} > \varepsilon_1 > \varepsilon_{2y}$, light propagating from the isotropic medium to the anisotropic one has degenerate Fresnel coefficients for $s$ and $p$ polarization ($r_s = r_p$) at a certain angle. This attribute originates from the occurrence of total internal reflection under the $s$-polarized light and the resultant $\pi$ phase shift in the Fresnel coefficients. We use the material parameters given as $\varepsilon_1 = 2, \varepsilon_{2x} = \varepsilon_{2z} = 4.1, \varepsilon_{2y} = 1$. In such a case, $r_s = r_p$ is satisfied at a certain angle defined as $\theta_0$ (Fig. \ref{random}b and c). A feasible design of such a medium and details of the reflection at the boundary of an anisotropic medium can be found in Supporting Information.

Because the analytic formula of the shift is known explicitly only for a horizontally or vertically polarized incidence, the shift under arbitrarily polarized incidence is obtained by taking the average $y$ position of the reflected beam profile numerically. Thus, we first obtain the spatial distribution of the reflected Gaussian beam. By combining Eq. \ref{incidence}-\ref{hv2lr} and then applying a Taylor expansion to reflection coefficients
\begin{equation}
	r_{p,s} \approx r_{p,s}(\theta_i) + \frac{k_{Ix}}{k_0} \frac{\partial r_{p,s}}{\partial \theta_i},
\end{equation}
the field distribution of the reflected beam can be found in momentum space. Then the spatial distribution of the reflected beam can be obtained by applying a Fourier transform
\begin{equation}
	\tilde{\psi}^\pm_R (x_R, y_R, z_R) = \iint{dk_{Rx} dk_{Ry} \psi^\pm_R(k_{Rx},k_{Ry}) \psi_0 \exp \big(i(k_{Rx}x_R + k_{Ry}y_R + k_{Rz}z_R) \big)}.
\end{equation}
Each circularly polarized component of a reflected beam under a horizontally and vertically polarized incidence has a spatial profile of
\begin{align}
	\tilde{\psi}^\pm_{R,H} =& \frac{1}{\sqrt{2\pi} \omega_0} \frac{z_0}{z_0 + i z_R} \exp(-\frac{k_0}{2}  \frac{x_R^2+y_R^2}{z_0 + i z_R}) \notag \\
	&\times \Big[ r_p  - i \frac{x_R}{z_0 + i z_R} \dot{r_p} \mp \frac{y_R \cot{\theta_i}}{z_0 + i z_R} (r_p + r_s) \mp i \frac{x_R y_R \cot{\theta_i}}{(z_0 + i z_R)^2} (\dot{r_p} + \dot{r_s}) \Big] \exp(ik_r z_R), \notag \\
	\tilde{\psi}^\pm_{R,V} =& \frac{\pm i}{\sqrt{2\pi} \omega_0} \frac{z_0}{z_0 + i z_R} \exp(-\frac{k_0}{2}  \frac{x_R^2+y_R^2}{z_0 + i z_R}) \notag \\
	&\times \Big[ r_s  - i \frac{x_R}{z_0 + i z_R} \dot{r_s} \mp \frac{y_R \cot{\theta_i}}{z_0 + i z_R} (r_p + r_s) \mp i \frac{x_R y_R \cot{\theta_i}}{(z_0 + i z_R)^2} (\dot{r_p} + \dot{r_s}) \Big] \exp(ik_r z_R),
	\label{fieldHV}
\end{align}
where the second subscript corresponds to incident polarization, the dot notation indicates the first derivative with respect to $\theta_i$, and $z_0 = k_0 \omega_0^2/2$ is the Rayleigh length. These four field distributions correspond to the elements of a matrix that transforms the Jones vector of the $n$-th incident polarization to the $n$-th reflected beam profile, i.e.,
\begin{equation}
	\tilde{\psi}^{\pm(n)}_R = \tilde{\psi}^\pm_{R,H} \psi_I^{H(n)} + \tilde{\psi}^\pm_{R,V} \psi_I^{V(n)}.
	\label{psiR}
\end{equation}


Eq. \ref{psiR} provides the spatial profiles of the two circularly polarized components of the reflected beam for each of the given incident polarization. We conduct $N$ independent calculations with parameters given as: $N = 1000, w_0 = 10^3 \lambda, z_R = 10 \lambda$ where $\lambda$ is the wavelength. In the simulation, $\lambda$ is set as 600 nm but has no influence on the results. The shifts that each of the given incidence undergoes are obtained by calculating beam centroids of the reflected beam as
\begin{equation}
	\delta^{\pm(n)} = \frac{\expval{y_R}{\tilde{\psi}_R^{\pm(n)}}}{\bra{\tilde{\psi}_R^{\pm(n)}}\ket{\tilde{\psi}_R^{\pm(n)}}},
	\label{yaverage}
\end{equation}
and are plotted as black dots in Fig. \ref{random}d and e, after normalized by $\lambda$. Because the shift depends on the incident polarization, the scatter plots exhibit many different $\delta^{\pm(n)}/\lambda$ at a fixed $\theta_i$. However, at $\theta_0$ at which $r_s = r_p$ is satisfied and thus $\delta_H^\pm = \delta_V^\pm$, the values of $\delta^{\pm(n)}/\lambda$ all appear at the intersection point regardless of the incident polarization (Fig. \ref{random}d and e, yellow dashed lines). The convergence of the shift for all incident polarization is also confirmed by the zero standard deviation of $\delta^{\pm(n)}/\lambda$ at $\theta_0$ (Fig. \ref{random}f). The standard deviation is nonzero at all $\theta_i$ other than $\theta_0$. The large standard deviation near the Brewster angle results from the diverging shift under horizontally polarized incidence. In several consecutive Monte Carlo simulations in which the $N$ polarization states are randomly redistributed, the fine details of the scatter plot in Fig. \ref{random}d and e change, but the overall tendency remains unaltered. Besides $\theta_0$, there exists another intersection of $\delta^\pm_{H,V}/\lambda$ near 60$^\circ$, but the shift is incident-polarization-dependent because $\triangle_H \neq \triangle_V$ at this angle and hence $\psi_R^\pm$ are not eigenstates of $i\partial_{k_y}$.

\begin{figure}[h!] \centering
	\includegraphics[width = 0.7031\textwidth]{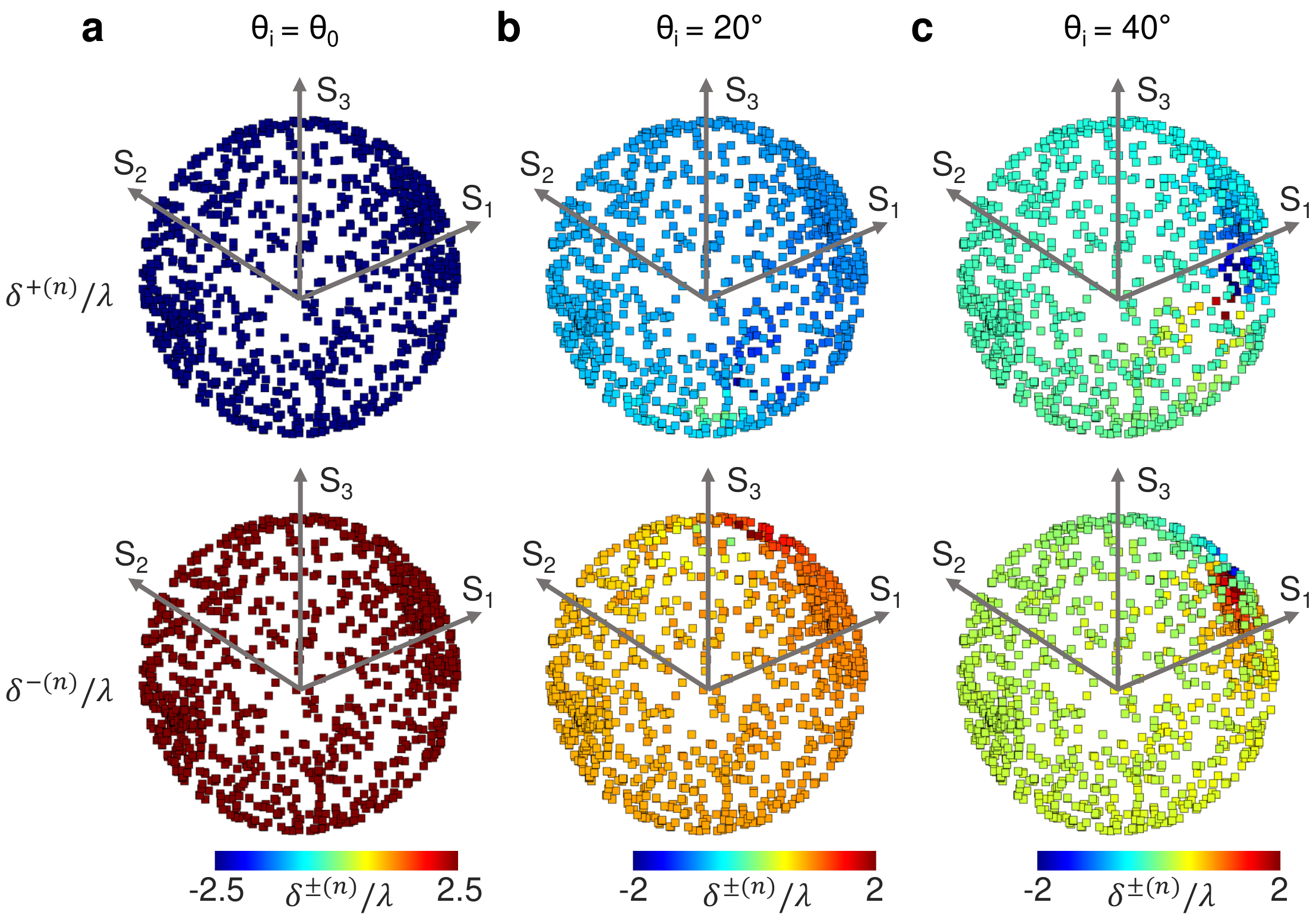} 
	\caption{Shift distribution represented in a Poincar\'e sphere when (a) $\theta_i = \theta_0$, (b) $\theta_i = 20^\circ \neq \theta_0$, and (c) $\theta_i = 40^\circ \neq \theta_0$. The location of a point on the sphere manifests the polarization states of the $n$-th incidence; the color represents $\delta^{\pm(n)}/\lambda$. Top and bottom corresponds to the shift of LCP and RCP components respectively. The boundary of the sphere is omitted for clarity.}
	\label{shift_sphere}
\end{figure}
Fig. \ref{random}d and e show the distributions of the shift for a given $\theta_i$, but do not provide the correspondence between the polarization state and the shift. Thus, we examine the distribution of $\delta^{\pm(n)}/\lambda$ in Poincar\'e sphere for three different $\theta_i$ (Fig. \ref{shift_sphere}). The location of a point on the sphere manifests the polarization states of the $n$-th incidence, and the color represents $\delta^{\pm(n)}/\lambda$. As confirmed by Fig. \ref{random}d-f, the shift under an arbitrarily polarized incidence are all degenerate and are also symmetrical for LCP and RCP components when $\theta_i = \theta_0$ (Fig. \ref{shift_sphere}a). However, at other $\theta_i = 20^\circ \neq \theta_0$, the shift depends on the incident polarization, showing variations of the shift over the sphere (Fig. \ref{shift_sphere}b). This tendency is more noticeable under circular and elliptical polarization than under linear polarization. In contrast, at $\theta_i = 40^\circ$, incidence that has Stokes parameters near $(1, 0, 0)$ produce shifts with large deviation (Fig. \ref{shift_sphere}c). This trend occurs because of the diverging shift under the horizontally polarized light near the Brewster angle. Furthermore, the shift distribution at $\theta_i \neq \theta_0$ shows that the splitting is not symmetrical ($\lvert \delta^{+(n)} \rvert \neq \lvert \delta^{-(n)} \rvert$) under arbitrarily polarized incidence if $r_s \neq r_p$ (Fig. \ref{shift_sphere}b and c).

\subsection{Numerical demonstration of the SHEL under unpolarized incidence}
\begin{figure}[h!] \centering
	\includegraphics[width = 0.7813\textwidth]{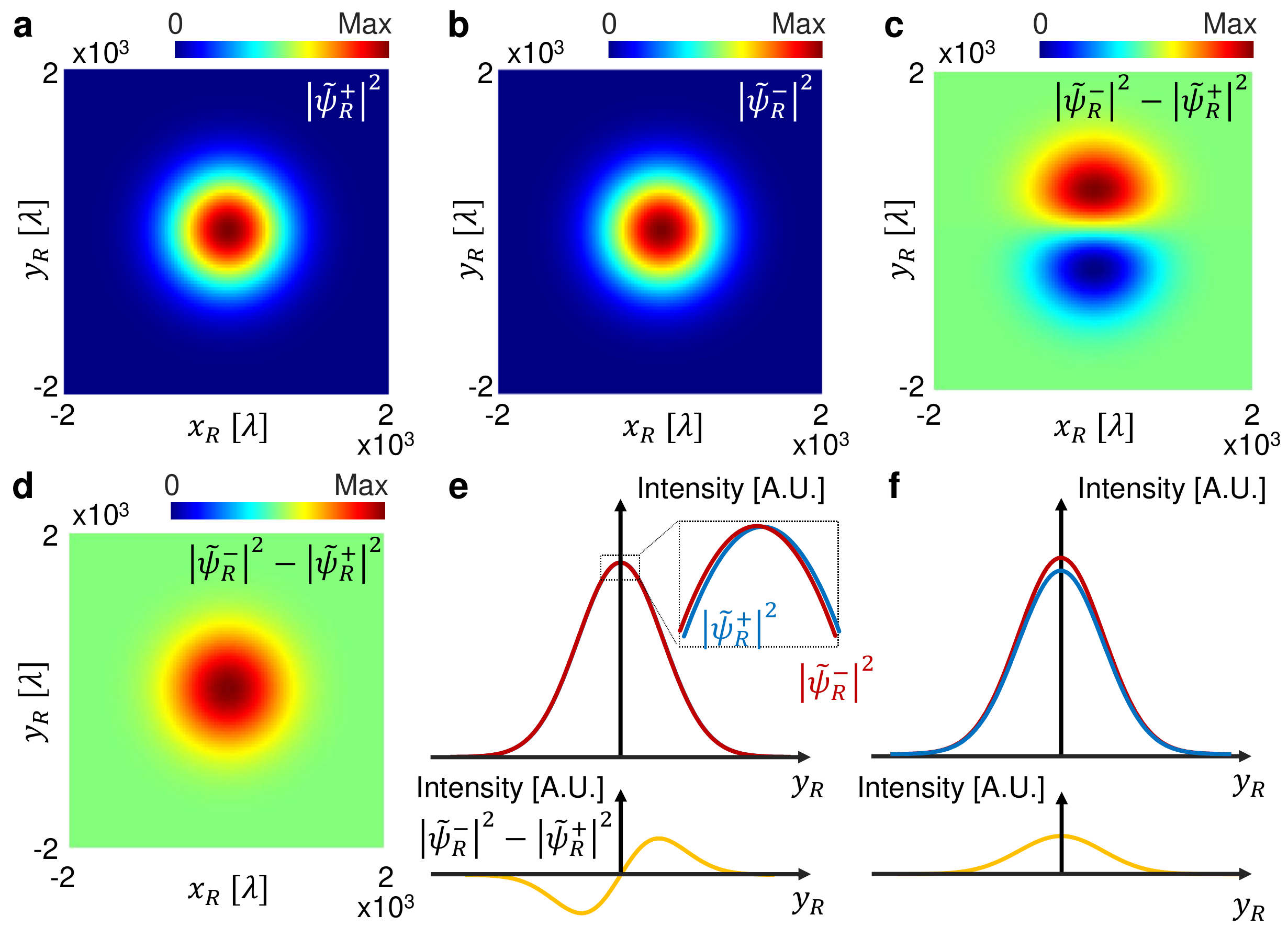} 
	\caption{Field profiles of the reflected Gaussian beam. Spatial distributions of (a) $\lvert \tilde{\psi}_R^+ \rvert ^2$, (b) $\lvert \tilde{\psi}_R^- \rvert ^2$, and (c) their difference $\lvert \tilde{\psi}_R^- \rvert ^2 - \lvert \tilde{\psi}_R^+ \rvert ^2$ at $\theta_0$. (d) Spatial distribution of $\lvert \tilde{\psi}_R^- \rvert ^2 - \lvert \tilde{\psi}_R^+ \rvert ^2$ at $\theta_i = 50^\circ \neq \theta_0$. (e, f) Intensity profiles in an arbitrary unit (A.U.) of the two circularly polarized components of the reflected beam (blue: $\lvert \tilde{\psi}_R^+ \rvert ^2$, red: $\lvert \tilde{\psi}_R^- \rvert ^2$) and their difference (yellow) along the transverse axis at $x_R = 0$ at (e) $\theta_0$ and (f) $\theta_i = 50^\circ \neq \theta_0$. For better visualization, $\lvert \tilde{\psi}_R^- \rvert ^2 - \lvert \tilde{\psi}_R^+ \rvert ^2$ are exaggerated in Fig. \ref{field}e and f.}
	\label{field}
\end{figure}
Because of the linearity of Eq. \ref{psiR}, the total reflected beam under unpolarized incidence can be obtained by taking a superposition of all $\tilde{\psi}^{\pm(n)}_R$ for a set of given incident polarizations: $\tilde{\psi}^\pm_R = \sum_{n = 1}^N \tilde{\psi}^{\pm(n)}_R/N$. For completeness, the intensity distributions of the two circularly polarized components of the total reflected beam and their difference are presented in Fig. \ref{field}. The splitting of the LCP and RCP is not readily apparent (Fig. \ref{field}a and b), but their intensity difference, which is proportional to $S_3$, shows the symmetric and spin-dependent shift along the transverse direction at $\theta_0$ (Fig. \ref{field}c). This SHEL under unpolarized incidence can be understood as a result of the superposition of the spin-dependent splittings under $N$ differently polarized incidences, in which the magnitude of the splittings are all degenerate. This field profile is obtained under unpolarized incidence but resembles that under a horizontally or vertically polarized incidence \cite{Bliokh:16, kim2019observation}.

In contrast, at $\theta_i = 50^\circ \neq \theta_0$, this splitting exhibits significantly distinct features (Fig. \ref{field}d) for the following reasons. Firstly, the shift is incident-polarization-dependent; the $N$ differently polarized incidences are split by different amounts of displacement, then superposed. Secondly, the $\lvert \tilde{\psi}_R^- \rvert ^2 - \lvert \tilde{\psi}_R^+ \rvert ^2$ distribution of a single sign is attributed to the asymmetrical splitting of intensity ($I_R^+ \neq I_R^-$) as a result of the phase difference between $r_s$ and $r_p$. If the shift is not large enough, the asymmetry in intensity leads to a single sign of $S_3$ distribution (Fig. \ref{field}d). For better understanding, the intensities along $x_R = 0$ are shown in Fig. \ref{field}e and f. Whereas the unpolarized incidence is split symmetrically into two circularly polarized components at $\theta_0$ (Fig. \ref{field}e), the SHEL at $\theta_i \neq \theta_0$ is asymmetrical in both shift and intensity (Fig. \ref{field}f).

\section{Conclusion}
The SHEL is known as a symmetrical splitting of a refracted or reflected beam into two circularly polarized beams along the transverse axis at an optical interface, but such symmetries only appear when the incidence is horizontally or vertically polarized. Here we show that the two circularly polarized components of the reflected beam undergo incident-polarization-independent and symmetrical splitting under an arbitrary incident polarization if $r_s = r_p$. The polarization independence of the shift is proved theoretically, then numerically using a Monte Carlo simulation. The intensities of the circularly polarized components of the reflected beams are symmetrical under unpolarized incidence, thereby appearing exactly the same as under horizontally or vertically polarized incidence. Lastly, an interface at which $r_s = r_p$ is suggested by adjoining isotropic and anisotropic dielectric media. In contrast to the previous research on the SHEL in which a well-defined polarization state is preliminary, the incident-polarization-independent SHEL can widen the boundaries and possible applications of the SHEL to cover unpolarized or ill-defined polarized sources.

\section{Method}
\subsection{Monte Carlo simulation}
For the random polarization states, the elements of the Jones vector of the $n$-th incidence are defined as $\psi_I^{H,V(n)} = \lvert \psi_I^{H,V(n)} \rvert \exp(i \phi^{H,V(n)})$ where the magnitude $\lvert \psi_I^{H,V(n)} \rvert$ and the phase $\phi^{H,V(n)}$ are real scalars that are randomly chosen in $(0, 1)$ and $(-\pi, \pi)$ respectively for $n \in \{1, ..., N\}$. Then the Jones vector $\begin{pmatrix} \psi_I^{H(n)} & \psi_I^{V(n)} \end{pmatrix}^T$ is normalized by its magnitude $\sqrt{\lvert \psi_I^{H(n)} \rvert^2 + \lvert \psi_I^{V(n)} \rvert^2}$. 

\subsection{Reflection coefficients at an interface between isotropic and anisotropic media}
At an interface between the isotropic (medium 1, permittivity $\varepsilon_1$) and anisotropic (medium 2, permittivity $\varepsilon_2 = \text{diag}(\varepsilon_{2x}, \varepsilon_{2y}, \varepsilon_{2z})$) media, the Fresnel reflection coefficients can be obtained by solving Maxwell's equations:
\begin{align}
	r_s =& \frac{\sqrt{\varepsilon_1 - \beta^2} - \sqrt{\varepsilon_{2y}- \beta^2}}{\sqrt{\varepsilon_1 + \beta^2} + \sqrt{\varepsilon_{2y} - \beta^2}}, \notag \\
	r_p =& \frac{\sqrt{\varepsilon_1 - \beta^2} /\varepsilon_1- \sqrt{\varepsilon_{2x} - \beta^2\varepsilon_{2x}/\varepsilon_{2z}}/\varepsilon_{2x}}{\sqrt{\varepsilon_1 - \beta^2} /\varepsilon_1+ \sqrt{\varepsilon_{2x}-\beta^2 \varepsilon_{2x}/\varepsilon_{2z}}/\varepsilon_{2x}}
	\label{Fresnel}
\end{align}
where $\beta = \sqrt{\varepsilon_1}\sin\theta_i$ is the propagation constant. 




\newpage
\bibliographystyle{custom}
\bibliography{reference}

\end{document}